\documentclass[aps,prb,reprint,twocolumn, superscriptaddress]{revtex4-1}
\usepackage{bbold}
\usepackage{amsmath,amsfonts,amsmath,mathtools,bbm,bm}
\usepackage{graphicx}
\usepackage{braket}
\usepackage[colorlinks,linkcolor=blue,urlcolor=blue,citecolor=blue]{hyperref}
\usepackage[all]{hypcap}

\usepackage{blindtext}
\usepackage{enumitem}
\usepackage{xcolor}
\usepackage[normalem]{ulem}

\setcitestyle{numbers,square}

\def\({\left (}
\def\){\right )}

\graphicspath{{Figure/PNG/}{Figure/PDF/}{Figure/EPS/}{Figure/TEX/}{Figure/}}

\newcommand\mydots{\hbox to 1em{.\hss.\hss.}}

\begin{document}

\title{Eigenstate Thermalization Hypothesis and Free Probability}

\author{Silvia Pappalardi}
\email{silvia.pappalardi@phys.ens.fr}
\affiliation{Laboratoire de Physique de l’\'Ecole Normale Sup\'erieure, ENS, Universit\'e PSL, CNRS, Sorbonne Universit\'e, Universit\'e de Paris, F-75005 Paris, France}

\author{Laura Foini}
\affiliation{IPhT, CNRS, CEA, Universit\'e Paris Saclay, 91191 Gif-sur-Yvette, France}

\author{Jorge  Kurchan}
\affiliation{Laboratoire de Physique de l’\'Ecole Normale Sup\'erieure, ENS, Universit\'e PSL, CNRS, Sorbonne Universit\'e, Universit\'e de Paris, F-75005 Paris, France}

\date{\today}

\begin{abstract}
Quantum thermalization is well understood via the Eigenstate Thermalization Hypothesis (ETH). The general form of ETH, describing all the relevant correlations of matrix elements, may be derived on the basis of a `typicality' argument of invariance with respect to local rotations involving nearby energy levels. In this work, we uncover the close relation between this perspective on ETH and Free Probability theory, as applied to a thermal ensemble or an energy shell. 
This mathematical framework allows one to reduce in a straightforward way higher-order correlation functions to a decomposition given by minimal blocks, identified as free cumulants, for which we give an explicit formula. 
This perspective naturally incorporates the consistency property that local functions of ETH operators also satisfy ETH. 
The present results uncover a direct connection between the Eigenstate Thermalization Hypothesis and the structure of Free Probability, widening considerably the latter's scope and highlighting its relevance to quantum thermalization.
\end{abstract}

\maketitle

{\bf{Introduction}} - 
The current framework for understanding the emergence of thermal equilibrium in isolated quantum systems goes under the name of the \emph{Eigenstate Thermalization Hypothesis} (ETH).
 Early works of Berry \cite{berry1977regular}, Deutsch \cite{deutsch1991quantum} and Srednicki \cite{srednicki1994chaos} recognized the importance of understanding the eigenstates of chaotic systems a pseudo-random vectors that encode microcanonical ensembles. 
Inspired by Random Matrix Theory (RMT), ETH was then fully established by Srednicki in Ref.\cite{srednicki1999approach}, incorporating some additional structure required to account for non-trivial temperature or time dependences. 
See Ref.\cite{dalessio2016from} for a review.
According to ETH, the matrix elements of local observables $A$ in the energy eigenbasis $H|E_i\rangle = E_i |E_i\rangle$ are pseudo-random numbers, whose statistical properties are smooth thermodynamic functions. In the original formulation, the average and variance read
\begin{equation}
\label{ETH}
    \overline{A_{ii}} = \mathcal A(E_i) \ , \quad \overline{A_{ij}A_{ji}}= F^{(2)}_{E_{ij}^+}(\omega_{ij}) e^{-S(E_{ij}^+)} \quad \text{for}\quad i\neq j\ ,
\end{equation}
where $E^+_{ij}=(E_i+E_j)/2$, $\omega_{ij}=E_i-E_j$ and $S(E)$ is the thermodynamic entropy at energy $E$. 
While $\mathcal A$ represents the microcanonical expectation value of $A$,  $F_E^{(2)}(\omega)$ depends implicitly on the observable $A$ ($|f_A(E, \omega)|^2$ with the standard notations \cite{dalessio2016from}) and it is associated to correlations on energy shell. 
In this paper, we will refer to them as \emph{on-shell correlations}. 
The ETH assumptions \eqref{ETH} allow one to fully describe the local relaxation of observables to thermal equilibrium as well as to characterize two-time dynamical correlation functions \cite{khatami2013fluctuation, dalessio2016from}. 
Since its formulation, ETH has motivated a considerable body of numerical \cite{rigol2008thermalization, rigol2010quantum, polkovnikov2011colloquium} and analytical work \cite{polkovnikov2011colloquium, anza2018eigenstate, jin2020equilibration,bauer2020universal}, also in relation to quantum entanglement \cite{vidmar2017entanglement, murthy2019structure, brenes2020multipartite}. Despite this progress, the precise relation between ETH and RMT is currently the focus of a large debate.
\vspace{.09cm}

With the recent explosion in activity imported from the String Theory community 
that revolutionized the field of Quantum Chaos \cite{maldacena2016bound},
the question of how ETH applies to multi-point correlations (as the out-of-time order correlators  OTOC) came to the forefront.  
Higher order correlators are important to several areas of many-body physics: from quantum information scrambling (through OTOCs) \cite{hosur2016chaos}, to dynamic-heterogeneity effects (through the fluctuation of correlations) \cite{berthier2011dynamical} or in pump-probe experiments (through three point functions) \cite{lloyd20212021}.
It became clear that a more general version of ETH had to be introduced, encompassing correlations between matrix elements hitherto neglected in the standard framework.  
In order to compute such correlations of $q>2$ times,  Ref.\cite{foini2019eigenstate} 
formulated an extension of Eq. (\ref{ETH}) based on {\em typicality} arguments \cite{goldstein2010long,goldstein2010normal,goldstein2006canonical,reimann2010canonical,reimann2016typical}, as  applied to  small rotations of nearby energy levels. 
The existence of matrix elements correlations on top of Eq.\eqref{ETH}, recently confirmed numerically \cite{brenes2021out, wang2021eigenstate}, has motivated discussions on the finer structure of the ETH beyond Gaussian RMT \cite{prosen1994statistical, dymarsky2018bound, chan2019eigenstate, murthy2019bounds, richter2020eigenstate}.\\
This perspective offered some understanding of the finite contribution of different matrix elements functions 
from a diagrammatic approach, although it did not provide an efficient calculational tool for the various terms, leaving unknown the general structure of multi-time correlation functions. 

\vspace{.09cm}

In this paper, we characterize this structure by identifying its intimate relation between the general form of the ETH and \emph{Free Probability theory}.
The latter can be thought of as the generalization of classical probability to non-commutative random variables, where the concept of 
``{freeness}'' extends the one of  ``independence''.
Introduced by Voiculescu 
in connection to the theory of operator algebras \cite{voiculescu1985symmetries}, Free Probability theory turned out to have important links with several branches of mathematics and physics \cite{ebrahimi2016combinatorics, morampudi2019many, bellitti2019hamiltonian},
such as RMT \cite{mingo2017free} and combinatorics. 
We are here interested in the combinatorial aspects of Free Probability,
based on free cumulants and 
non-crossing partitions \cite{speicher1997free}. \\
Using these tools, we show that the higher-order correlation functions of generic physical systems are determined by basic quantities: the thermal free cumulants, thus providing a sort of generalized Wick theorem.
 Our methodology is to use the properties of the ETH matrix elements and their diagrammatic description to link them with the free probability mathematical structure. 
 We will first recall the derivation of the general form of ETH, based on invariance with respect to local rotations of nearby energy levels. By discussing the ETH diagrams relevant to correlation functions, we will show that they are in one-to-one correspondence to non-crossing partitions. 
 Our main result is an explicit expression for the thermal free cumulants 
 in terms of sums of the matrix elements over non-repeated indices: simple loops in the diagrammatic language discussed in \cite{foini2019eigenstate}. 
The thermal free cumulants are hence linked to the Fourier transform of the ETH on-shell correlation of order $q$. As a byproduct, free probability allows us to deduce bounds on the behaviour of on-shell correlations in the frequency domain.\\

\begin{figure}[t]
	\centering
	\includegraphics[width=.9 \linewidth]{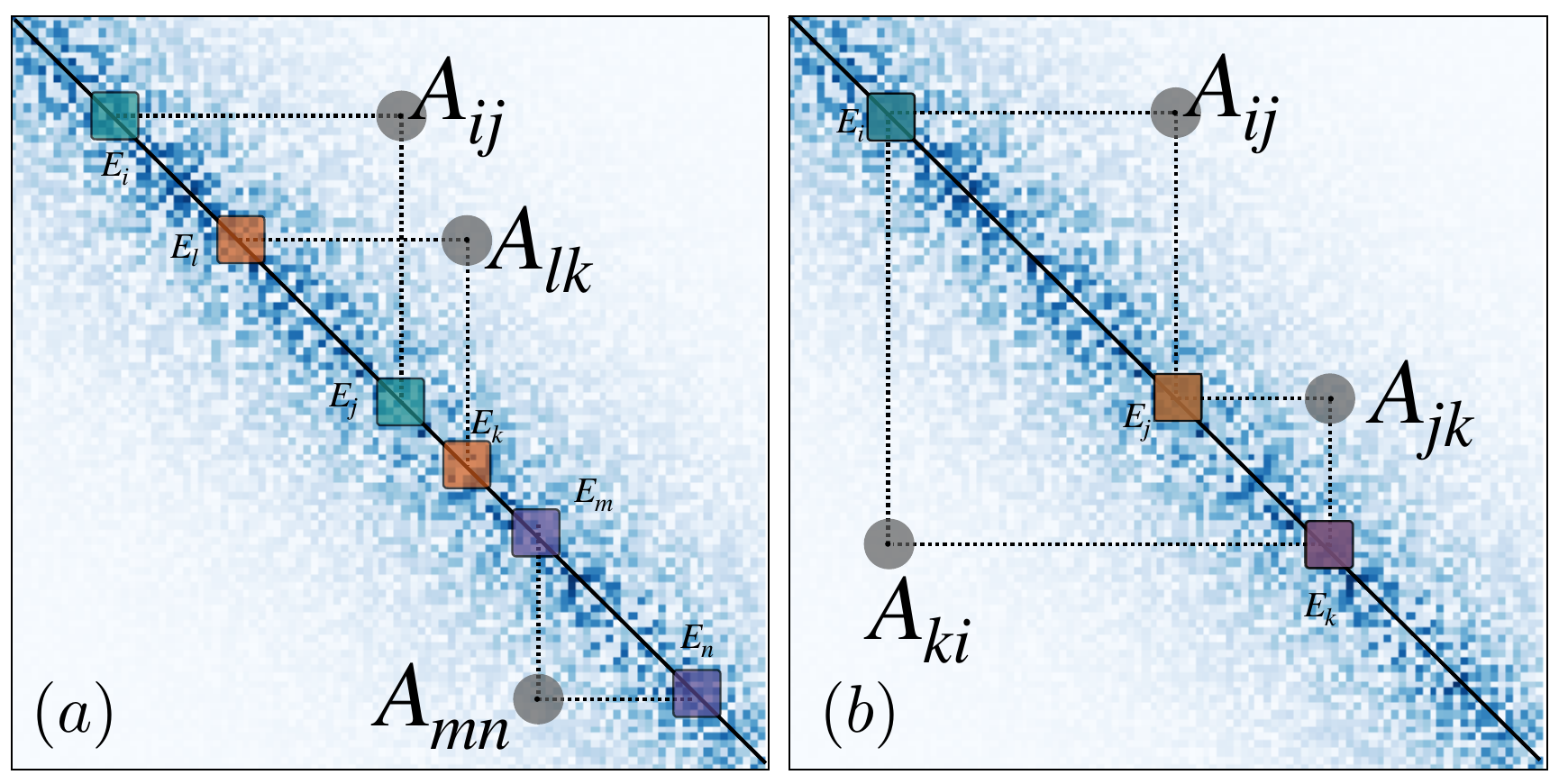}
	\caption{Impact of the local rotational invariance of $A_{ij}$ on the correlations between three matrix elements. The operator $A$ in the energy eigenbasis is depicted as a random matrix with a band structure. To each matrix element $A_{ij}$ is associated with a ``small'' $U$ (box on the diagonal) which acts as a pseudo-random unitary matrix. Matrix elements with different indices (a) are characterized by different $U$ and their average vanishes. When the indices are repeated on a loop (b) the $U$ appear in pairs and yield a finite result.}
	\label{fig:eth}
\end{figure}

{ \bf{General ETH - }}
The ETH in its enlarged formulation was discussed in Ref.\cite{foini2019eigenstate} to compute correlation functions of order $q$ depending on $\vec{t}=(t_1,t_2,\dots,t_{q-1})$ times. The latter can be written in terms of the product of $q$ matrix elements. The ETH amounts in the following ansatz: the average of products with distinct indices $i_1\neq i_2 \mydots \neq i_q$ reads
\begin{equation}
    \label{ETHq}
    \overline{A_{i_1i_2}A_{i_2i_3}\mydots A_{i_{q}i_1}} = e^{-(q-1)S(E^+)} F_{E^+}^{(q)}(\omega_{i_1i_2}, \mydots, \omega_{i_{q-1}i_q}) 
\end{equation}
and with repeated indices it factorizes in the large $N$ limit
\begin{align}
\label{ETH_conta}
 \overline{A_{i_1i_2}\mydots A_{i_{k-1}i_1}A_{i_1i_{k+1}}\mydots A_{i_{q}i_1}}& \\
 = & \overline{A_{i_1i_2}\mydots A_{i_{k-1}i_1}} \; 
\overline{A_{i_1i_{k+1}}\mydots A_{i_{q}i_1}}    \ .   \nonumber 
\end{align}
In Eq.\eqref{ETHq}, $E^+=(E_{i_1}+\mydots +E_{i_q})/q$ is the average energy, $\vec 
\omega = (\omega_{i_1i_2}, \dots, \omega_{i_{q-1}i_q})$ with $\omega_{ij}=E_i-E_j$ are $q-1$ energy differences and $F_{E^+}^{(q)}(\vec \omega)$ is a smooth function of the energy density $E^+/N$ and $\vec \omega$. Thanks to the explicit entropic factor, $F^{(q)}_E(\vec \omega)$ is of order one and thus contains Eq.\eqref{ETH} as a particular case for $q=1, 2$.
We will refer to  $F^{(q)}_E(\vec \omega)$ as the \emph{on-energy shell correlations of order $q$.} 
This generalization of ETH, which is necessary if matrix elements are considered to be not independent, implies that correlation functions at order $q$ contain new information that is not in principle encoded in lower moments.

The ETH ansatz in Eq.\eqref{ETHq} can be derived using typicality arguments. The central idea is to use local invariance of the $A_{ij}$, stemming from small rotations that involve only nearby energy levels. 
The matrix elements are evaluated by substituting the operator with a ``locally'' rotated one (see Fig. \ref{fig:eth}) $A^u = U A U^\dagger$, i.e. $A_{ij} = \sum_{\bar i \bar j}U_{i \bar i} A_{\bar i \bar j} U^{*}_{\bar j j}$, with  $U_{i\bar i} = \langle E_i|E_ {\bar i}\rangle $ and $|E_ {\bar i}\rangle$ are the eigenstates of a slightly perturbed Hamiltonian \cite{deutsch1991quantum}.
By looking at a sufficiently small energy range around $E_i$, the  $U_{i \bar i}$ can be thought as a pseudorandom unitary matrix. This is in analogy to Berry's conjecture, stating that the overlaps of chaotic eigenstates with a generic basis can be thought as random Gaussian numbers.
The size of this matrix has to be ``small'' 
to keep intact the energy band structure of $A_{ij}$, but it contains many level spacings. Hence the matrix elements are treated as belonging to an ensemble of \emph{local rotational invariances}. By averaging over $U$, one can immediately deduce the finite contributions to any product of matrix elements. Averages are non-zero only if the matrices $U$  appear at least twice. For example, for $\overline{A_{ij}A_{lk}}$ the only finite contribution comes from $\overline{A_{ij}A_{ji}}$ leading to Eq.\eqref{ETH}. 
In the same way, finite products of $q$ matrix elements necessarily have to be in a loop (see Fig.\ref{fig:eth} for the pictorial example with $q=3$). When the indices are different, this leads to Eq.\eqref{ETHq}. This scenario, complemented with some entropic arguments, results also in the factorization of Eq.\eqref{ETH_conta}, see Ref.\cite{foini2019eigenstate}.

As a consequence, we remark that this approach also accounts for the validity of the ETH ansatz between different operators, e.g. $\overline{A_{ij}B_{ji}}=  F^{(2)}_{E^+, AB}(\omega)e^{-S(E^+)}$, where we make explicit the dependence on the operators.  Clearly, the ensemble defined by $U$ is the same for $A$ and $B$, since they come only from changes in $H$. Hence, the above argument can be applied to any set of local observables. \\

\begin{figure*}[t]
	\centering
	\includegraphics[width=1 \linewidth]{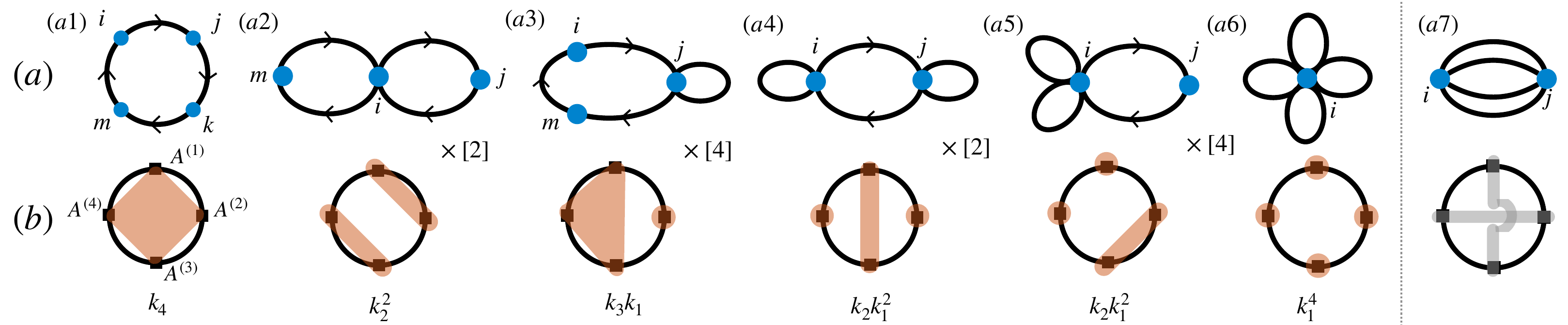}
	\caption{ETH diagrams (a) and non-crossing partitions (b) for $q=4$. (a1-6) Loop and cactus diagrams that contribute to ETH correlators. The arrow indicates the presence of a time dependence. With $\times [n]$ we indicate that there are $n$ cyclic permutations. (a7) Non-cactus diagram. (b1-6) Non-crossing partitions for $q=4$.  Each of the blocks contributes with a free cumulant $k_{n}$, where $n$ is the number of $n$ points in that partition. For completeness, we also represent the crossing partition after the dashed line. }
	\label{fig:cumufree}
\end{figure*}

{\bf{Computing expectations: diagrammatic expansion -}}
The ETH ansatz in Eq.\eqref{ETHq} allows one to compute multi-time thermal correlation functions of the form
\begin{eqnarray}
    \label{eq:Sq}
    S_\beta^{(q)} (\vec t) = \langle A(t_1) A(t_2) \dots A(t_{q-1}) A(0) \rangle_\beta
\end{eqnarray}
where $\langle \bullet \rangle_\beta = \text{Tr}( \rho \, \bullet )$ and $\rho = e^{-\beta H}/Z$ with ${Z = \text{Tr}(e^{-\beta H})}$. Here, $A(t) = e^{i H t} A e^{- i H t}$ is the observable at time $t$ in the Heisenberg representation ($\hbar=1$). 

As introduced in Ref.\cite{foini2019eigenstate}, one can determine the contribution of $\overline{A_{i_1i_2}\dots A_{i_{q}i_1}}$ to the thermal correlation $S_\beta^{(q)} (\vec t)$ in a \emph{diagrammatic fashion}, see Fig.\ref{fig:cumufree}a for $q=4$.
Let us briefly illustrate how to understand pictorially such ETH diagrams. Products as $A_{i_1i_2}\dots A_{i_{q}i_1}$ are represented on a loop with $q$ points. The matrix elements $A_{ij}$ live on the oriented edge connecting two points $i$ and $j$.  The arrows keep track of non-trivial time dependencies. The different diagrams correspond to all the different ways one can contract the indices, i.e. identify two points. 
Such diagrams are classified as:

\indent $\bullet$    \emph{loops}:  all distinct vertices lie on a single closed circle (e.g. $A_{ij}A_{jk}A_{km}A_{mi}$ in Fig.\ref{fig:cumufree}a1). Each loop with $n$ vertices contributes with $\propto F^{(n)}e^{-(n-1)S}$;

\indent $\bullet$ \emph{cactus diagrams}: trees of loops are joined to one another at single vertex (e.g. $A_{ij}A_{ji}A_{im}A_{mi}$ in Fig.\ref{fig:cumufree}a2). Cactus with $p$ leaves contribute with $p$ products of the associated $F$. The example of a two-leaf cactus in Fig.\ref{fig:cumufree}a2 contributes with $\propto (F^{(2)}e^{-S})^2$.

The \emph{non-cactus} diagrams (e.g. $A_{ij}A_{ji}A_{ij}A_{ji}$ in Fig.\ref{fig:cumufree}a7) have a further constraint to indices with respect to the other diagrams, their contribution is subleading for the correlation functions, as we now argue.

The thermal correlation  $S^{(q)}_\beta(\vec t)$ in Eq.\eqref{eq:Sq} is given by the sum over all indices (and correspondingly all the diagrams) with the proper Boltzmann weight $e^{-\beta E_i}/Z$. The ETH ansatz \eqref{ETHq}-\eqref{ETH_conta} results in two main outcomes:
1) all summations of elements with repeated indices (cactus diagrams) factorize on results computed at the thermal energy $E_\beta = \langle H\rangle_\beta$. This means that leaves may be severed.
2) The contribution of non-cactus diagrams is exponentially small with respect to the other terms. These properties follow from the smoothness of the ETH functions and the proper entropic counting.
As an explicit example of 1), one can compute the diagram (a6) of Fig.\ref{fig:cumufree}, i.e.
\begin{align}
    \label{eq:saddleA4}
    \frac 1Z \sum_i {e^{-\beta E_i}} \overline{A_{ii}}^4 & = \frac 1Z \int dE {e^{-\beta E + S(E)}} \mathcal A^4(E) = (S^{(1)}_\beta)^4
\end{align}    
where we have substituted the ETH ansatz \eqref{ETH}, summations with integrals $\sum_i = \int dE e^{S(E)}$ and performed the integral in $E$ via saddle point technique, which fixes the energy by the thermodynamic condition $\beta = S'(E_\beta)$ and yields $\mathcal A^4(E_\beta) = (S^{(1)}_\beta)^4$. On the other hand, performing the same steps on the non-cactus diagram Fig.\ref{fig:cumufree}(a7) and expanding the entropies \cite{dalessio2016from} leads to
\begin{align}
        \frac 1Z & \sum_{i\neq j} {e^{-\beta E_i}} \overline{|A_{ij}|^2}\, \overline{|A_{ij}|^2} 
    \\ 
    & 
    = \frac 1Z \int dE^+d\omega  e^{-\beta E^+}  e^{-\beta \omega/2}
     \left (F^{(2)}_{E^+}(\omega) \right )^2 \sim \mathcal O(e^{-N}) \ .    \nonumber
\end{align}
In this paper, we rationalize that cactus diagrams correspond in fact to the non-crossing partitions that play a role in free probability theory. \\

{\bf{Hints of free probability theory - }}
We are interested in the combinatorial aspects of free probability, which are based on non-crossing partitions and free cumulants, as developed by Speicher \cite{speicher1997free}. 
A partition of a set $\{1, \dots q\}$ is a decomposition in blocks that do not overlap and whose union is the whole set.  Partitions in which blocks do not ``cross'' are called \emph{non-crossing partitions}. The set of all non-crossing partitions of $\{1, \dots q\}$ is denoted $NC(q)$. See the example in Fig.\ref{fig:cumufree}b for the partitions with $q=4$, with $\times [n]$ we denote the $n$ cycling permutations. There are 14 non-crossing partitions of $q=4$ elements, and only one is crossing, i.e. Fig.\ref{fig:cumufree}b.a7. \\
Non-crossing partitions appear in the definition of free cumulants. Consider some normalized $\phi$ (i.e. $\phi(1)=1$), for example $\phi(\bullet) = \text{Tr}(\bullet )/D$ for large $D\times D$ random matrices. The \emph{free cumulants} $k_q$ are defined implicitly from the moment-cumulant formula, stating that the moments of variables $A^{(i)}$ read
\begin{equation}
    \label{eq_free_cumu_def}
    \phi(A^{(1)} \dots A^{(q)} ) = \sum_{\pi \in NC(q)} k_{\pi} (A^{(1)} \dots A^{(q)} )  \ ,
\end{equation}
where $\pi$ is a non-crossing partition.
Here $k_{\pi}$ is a  product of cumulants, one term for each block of $\pi$. For instance for $A^{(i)}=A$ $\forall i$ and $k_n(A ... A) = k_n$, one has $\phi (A) = k_1$, $\phi (A^2) = k_2+ k_1^2$, $\phi (A^3) = k_3+ 3\, k_1 k_2 + k_1^3$, and $\phi(A^4) = k_4 + 2 k^2_2 + 4 k_3 k_1 + 6 k_2k_1^2 + k_1^4$, i.e. Fig.\ref{fig:cumufree}b.
Classical cumulants are defined by a similar formula, where one sum over all the possible partitions and not only on the non-crossing ones. Notably, the relation between moments and cumulants differs in classical and free probability only for $n\geq 4$.
Another interesting property is that the third and higher-order free cumulants of a Gaussian random matrix $A$ vanish, i.e. $k_{q\geq 3}(A...A)=0$ ( as for the classical cumulants of standard Gaussian random variables).
At this level, Eq.\eqref{eq_free_cumu_def} is a simple implicit definition of free-cumulants in terms of moments, e.g. $k_1 = \phi (A)$, $k_2 = \phi (A^2) - \phi (A)^2$, etc..\\

{\bf{ETH in these words -}}
The relation between ETH and free cumulants is thus clear: cactus diagrams correspond to non-crossing partitions, on the set identified by the matrices $A(t_i)$.  Likewise, non-cactus diagrams, not being associated with any non-crossing partition, do not count and they are related to crossing partitions \cite{SM}.
As we argued, they do not contribute to the thermal moments. 
Analogously to Eq.\eqref{eq_free_cumu_def}, we introduce an implicit definition of the \emph{thermal free cumulants}
\begin{equation}
    S_\beta^{(q)}(\vec t) = \sum_{\pi \in NC(q)} k^{\beta}_\pi(\, A(t_1) A(t_2) \dots A(0)) \ .
\end{equation}
Also here, $k^{\beta}_\pi$ is a product of free cumulants, one for each block of the partition $\pi$, as $k^{\beta}_{n}$ associated to $n$ operators. 
Note that the cumulant $k^\beta_{n}(\vec t)$ depends on the order in which we consider different operators, and we make this implicit in its time dependence.
\emph{The ETH ansatz \eqref{ETHq} is then the precise statement that the thermal free cumulants, may be substituted, for the purposes of computing time correlations, by sums  as
\begin{align}
    \label{free_cumuETH}
    k^\beta_{q}(\vec t) & = k^{ETH}_{q}(\vec t) 
    \\
    & = \frac 1Z \sum_{i_1\neq i_2 \neq i_{q}} e^{-\beta E_{i_1}} 
    A(t_1)_{i_1i_2}A(t_2)_{i_2i_3}\dots A(0)_{i_{q}i_1} \ .        \nonumber
\end{align} where all indices are different}.
In other words, free cumulants are simply given by the loop diagrams.
This follows from two properties of the ETH ansatz discussed above: a) that cactus diagrams factorize and b) that only cactus diagrams (non-crossing partitions) matter. This first point is almost trivial for $q=2$, where it is well known that via ETH one can compute
\begin{align}
    k^\beta_2(t) & \equiv
    S_\beta^{(2)}(t)-[S_\beta^{(1)}]^2  = 
    \langle A(t) A(0) \rangle_\beta - \langle A\rangle^2_\beta
    \\ &
    =\frac 1Z \sum_{i\neq j} {e^{-\beta E_i }}|A_{ij}|^2 e^{i(E_i-E_j)t} = k^{ETH}_2(t)    
\end{align}
where one uses that the diagonal ETH matrix element is a smooth function of energy and therefore $\sum_i e^{-\beta E_i }A_{ii}^2 \simeq \langle A\rangle_\beta^2$ by saddle point integral, as in Eq.\eqref{eq:saddleA4}. One can show that this factorization holds at all orders.  For instance, for $q=4$ fixing $k_1(A)=\langle A\rangle_\beta=0$, we obtain
\begin{align}
\begin{split}
    \label{eq_mom4}
    \langle A(t_1) & A(t_2) A(t_3) A(0) \rangle _\beta
     = k^{\beta}_4(t_1, t_2, t_3) \\
    & + k^{\beta}_2(t_1-t_2)\, k^{\beta}_2(t_3)
    + k^{\beta}_2(t_2-t_3) \, k_2^{\beta}( t_1 )   \ ,    
\end{split}
\end{align}
where $k_4^\beta$ is the term coming from the simple loop in Fig.\ref{fig:cumufree}a1 and encodes all the correlations beyond gaussian \cite{murthy2019bounds, brenes2021out}. This expression now immediately follows from free probability expression (the diagrams (b1-b2) of Fig.\ref{fig:cumufree}), while it would require in principle a lengthy calculation \cite{SM}.\\

Also in rotationally invariant random matrix ensembles, free cumulants are associated with diagrams with all distinct indices \cite{maillard2019high} and one can show that only the cactus diagrams matter.

Let us see how the structure of Free Probability leads to further results in the ETH context. First of all, it incorporates the consistency condition that products of operators obeying ETH shall obey ETH \cite{srednicki1996thermal, srednicki1999approach}. This can be checked directly from the Free Probability non-crossing partitions, see \cite{SM}.
 Secondly, we are led to ask questions such as the value of
    $ k_q^{ETH}(0)=  \frac 1Z \sum_{i_1\neq i_2 \neq i_{q}} e^{-\beta E_{i_1}} 
    A_{i_1i_2}A_{i_2i_3}\dots A_{i_{q}i_1} $.
Free Probability offers powerful computational tools to study such correlations via the generating functions \cite{bun2017cleaning}.  
Given the Stieltjes transform $G_\beta(z) = \text{Tr}(\rho \frac 1{z-A})$, related to the  generating function of thermal moments, one can study the so-called $R$-transform 
\begin{equation}
    R_\beta(w)  \equiv G^{-1}_\beta(w) - \frac 1w = \sum_{q=1} k_q^{\beta}(0) w^{q-1} \ ,
\end{equation}
that is always the generating function of equal-times free cumulants. In the case of ETH, $k^\beta_q(0) = k_q^{ETH}(0)$ and it generalizes the result of fully rotational invariant random matrices \cite{maillard2019high}. 
Finally, Free Probability offers the tools -- via the free cumulants --  to pinpoint and characterize the non-Gaussian aspects of
ETH \cite{brenes2021out, wang2021eigenstate, prosen1994statistical, dymarsky2018bound, chan2019eigenstate, murthy2019bounds, richter2020eigenstate}. 
 \\

{\bf{Free cumulants on-shell} -}
The thermal free cumulants defined in Eq.\eqref{free_cumuETH} admit an extremely appealing expression in terms of the ETH ansatz \eqref{ETHq}.  By standard manipulations \cite{SM}, one shows that 
\begin{equation}
    \label{free_thermal}
    k^\beta_q(\vec t) = \int d \vec \omega  F^{(q)}_{E_\beta}(\vec \omega) \, e^{i \vec \omega \cdot \vec t - \beta  \vec \omega \cdot \vec \ell_q} \ ,
\end{equation}
where we defined the thermal shift
$ {\vec{\ell}_q = \left (\frac {q-1}q,\frac {q-2}q, \dots, \frac 1q  \right )}$. 
This equation gives an important property: \emph{ETH $q$-th on-shell correlations are related to the Fourier transform of the thermal free cumulants $k^\beta_q$}
\begin{equation}
    \label{eq:generaKMS}
  k^\beta_q(\vec \omega) = F^{(q)}_{E_\beta}(\vec \omega) e^{- \beta \vec \omega \cdot \vec \ell_{q}}    \ .
\end{equation}
This is familiar for $q=2$, for which $k^\beta_2(\omega) =F^{(2)}_{E_\beta}(\omega) e^{- \beta \omega /2}$, which is the standard Kubo-Martin Schwinger (KMS) relation, leading to the fluctuation-dissipation theorem. 
The presence of this thermal shift -- which only depends on temperature and on the correlation function  order $q$ -- shall be interpreted as a generalized KMS condition, see Ref.\cite{bun2017cleaning}.  
Eq.\eqref{eq:generaKMS} naturally leads us to inspect the free cumulant expansion of the shifted correlator $\overline S^{(q)}_\beta(\vec t) \equiv S^{(q)}_\beta(\vec t - i \beta \vec \ell_q)$ given by
\begin{equation}
    \label{eq:shiftS}
    \overline S^{(q)}_\beta(\vec t) 
    = \text{Tr} \left ( \rho^{1/q} A(t_1) \rho^{1/q}  \dots A(t_{q-1}) \rho^{1/q} A(0) \right ) \ ,
\end{equation}
which corresponds to a \emph{regularized version} of $S_\beta$. 
One can look at the following connected correlation part of $\overline S^{(q)}_\beta$, i.e.
\begin{equation}
    \bar k_{q}(\vec t) =
    \frac 1Z \hspace{-.2cm}
    \sum_{i_1\neq \dots \neq i_{q}} 
    \hspace{-.3cm}
    e^{-\frac \beta q(E_{i_1}+\dots E_{i_q})} 
    A(t_1)_{i_1i_2}A(t_2)_{i_2i_3}\dots A(t_{q})_{i_{q}i_1} \ .
\end{equation}
Diagrammatically, it is associated with the loop with $q$ operators where the thermal weight $\rho^{1/q}$ is equally split. Nicely, its Fourier transform coincides with on shell correlations at the  energy $E_\beta$, i.e. \cite{foini2019eigenstate} $\label{eq:regulaYAY}
    \bar k_{q}^\beta(\vec \omega) = F^{(q)}_{E_\beta} (\vec \omega)$.
This allows accessing such correlations directly from the time-dependent correlation functions in time and by taking their Fourier transform.

We now recall that correlation between matrix elements with large energy differences should be small. This means that ETH correlation functions are usually expected to decay fast at large frequencies $\omega \gg 1$ as
\begin{equation}
\label{decay2}
    F^{(q)}_{E}(\omega)\sim e^{-|\omega|/\omega^{(q)}_{max}} \ .
\end{equation}
The relations between free cumulants and $F^{(q)}_E(\vec \omega)$ \eqref{eq:generaKMS} allow one to infer relevant properties of the latter. Using Eq.\eqref{free_thermal} and the fact that free cumulants at equal times shall be well defined, in \cite{SM}  we prove that on-shell correlations must decay at large frequencies in all directions at least as
\begin{equation}
    \label{eq:genera}
    F^{(q)}_{E_\beta}(\vec \omega) \sim \exp \left (-\beta \frac{q-1}{q} |\omega_i| \right ) 
\end{equation}
$\forall i=1, \dots, q-1$. This gives the bound $\omega^{(q-1)}_{max}\leq \frac{q-1}{q\beta}$, which generalizes the result for $q=2,4$ of  Ref.\cite{murthy2019bounds}.  
These kinds of constraints have been related to operator growth or to time scales of multi-time correlation functions (such as out-of-time order correlators) \cite{parker2019universal, murthy2019bounds, gu2022two,avdoshkin2020euclidean, cao2021statistical}, which have been proven to obey strict bounds \cite{maldacena2016bound, tsuji2018bound, pappalardi2022quantum}. \\

{\bf{Conclusions} -}
We have found that the ETH, when generalized to all multi-point correlations in the spirit as Berry, Deutsch and Srednicki, 
leads us directly to place it in the realm of Free Probability. This is a branch of mathematics where many developments have been made, and for which one may now turn to look for connections and analogies.

There is, however, a fundamental new element: the ensembles of matrices are not homogeneously full, but rather have a band structure, and a large, slowly varying diagonal.
This structure exists on a specific basis, the one where the Hamiltonian is diagonal and its eigenvalues are ordered.
The results are, likewise, always related to a specific energy shell, and not the matrix as a whole.  This is a distinguishing feature of using Free Probability within ETH to respect to standard RMT results. 
The moments that define equilibrium correlation functions are then more complicated objects than those of a standard rotationally-invariant Matrix Model. Nonetheless many results from these appear to generalize to the ETH setting and call for a rigorous understanding.

The ETH is at its most interesting when it fails, and integrability or many-body localization phenomena emerge. A more global understanding of ETH may then lead to a finer understanding of these effects.\\

\begin{acknowledgments}
This paper has been submitted simultaneously with ``Dynamics of Fluctuations in the Open Quantum SSEP and Free Probability'' by L. Hruza and D. Bernard \cite{hruza2022dynamics}, which discusses the appearance of free cumulants in stochastic transport models.  The occurrence of free probability in both problems has a similar origin:  the coarse-graining at microscopic either spatial or energy scales, and the unitary invariance at these microscopic scales.  Thus the use of free probability tools promises to be ubiquitous in chaotic or noisy many-body quantum systems. \\

We thank A. Polkovnikov for useful suggestions. SP thanks C. Malvenuto for discussions on non-crossing partitions.
SP and JK are supported by the Simons Foundation Grant No. 454943. This work is supported by `Investissements d'Avenir' LabEx PALM
(ANR-10-LABX-0039-PALM) (EquiDystant project, L. Foini). S. P. has received funding from the European Union’s Horizon Europe program under the Marie Sklodowska Curie Action VERMOUTH (Grant No. 101059865). 
\end{acknowledgments}

\bibliography{biblio}

------------------------------
\widetext
\clearpage
\begin{center}
\textbf{\large \centering Supplemental Material:\\ Eigenstate Thermalization Hypothesis and Free Probability   }
\end{center}

\setcounter{equation}{0}
\setcounter{section}{0}
\setcounter{figure}{0}
\setcounter{table}{0}
\setcounter{page}{1}
\renewcommand{\theequation}{S\arabic{equation}}
\setcounter{figure}{0}
\renewcommand{\thefigure}{S\arabic{figure}}
\renewcommand{\thepage}{S\arabic{page}}
\renewcommand{\thesection}{S\arabic{section}}
\renewcommand{\thetable}{S\arabic{table}}
\makeatletter

\renewcommand{\thesection}{\arabic{section}}
\renewcommand{\thesubsection}{\thesection.\arabic{subsection}}
\renewcommand{\thesubsubsection}{\thesubsection.\arabic{subsubsection}}

\newcommand{\nocontentsline}[3]{}
\newcommand{\tocless}[2]{\bgroup\let\addcontentsline=\nocontentsline#1{#2}\egroup}

In this Supplementary Material, we provide additional analysis and background calculations to support the results in the main text. In Sec.\ref{app:diagrams} and Sec.\ref{app:consistency} we discuss further properties of non-crossing partitions in relation to ETH, In Sec.\ref{app:freETH}, we derive the expression of the free cumulants within ETH.
In Sec.\ref{eq:large_omega} we discuss the large frequency dependence of on-shell correlations and prove Eq.\eqref{eq:genera} of the main text.
We conclude with Sec.\ref{sec_cumu-moment} with the detailed calculation of the moment-cumulant calculation for $q=4$.

\section{Cactus diagrams, non-crossing partitions and their duals}
\label{app:diagrams}

\begin{figure}[h]
	\centering
	\includegraphics[width=1\linewidth]{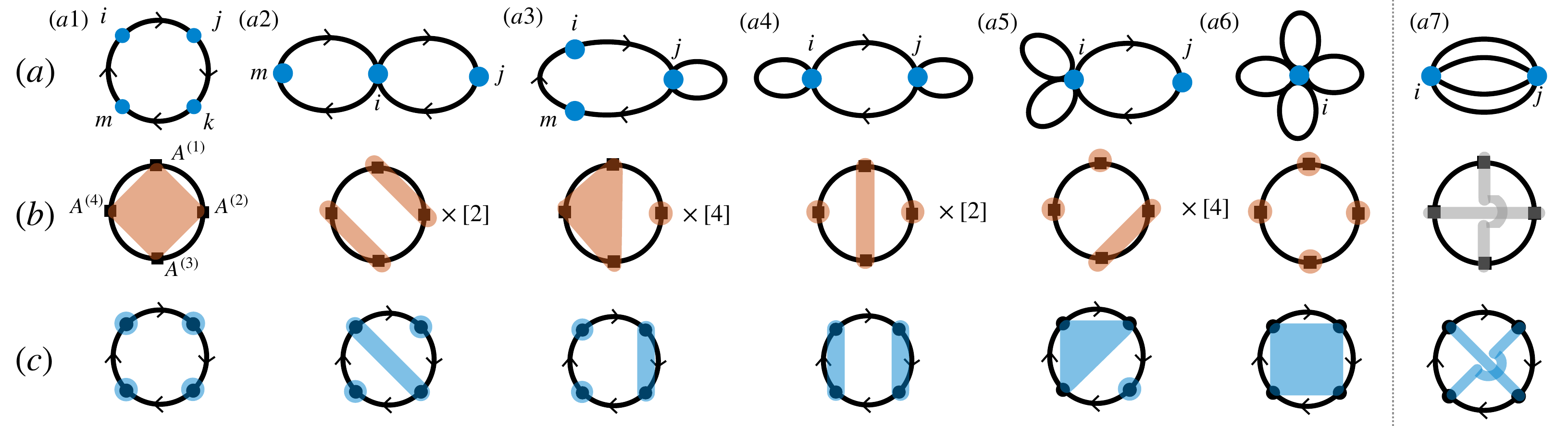}
	\label{fig:cumufree_ethDual}
	\caption{Relation between ETH diagrams (a), non-crossing partitions (b) and their dual (c) for $q=4$.}
	\label{fig:dual}
\end{figure}

Non-crossing partitions admit a complementary/dual representation, where the partition is identified on the vertices rather than on the links. 
This is illustrated in Fig.\ref{fig:dual} where we represent the example for $q=4$. The relation between the ETH diagrams (Fig.\ref{fig:dual}a) and the dual of the non-crossing partition Fig.\ref{fig:dual}c is thus particularly simple: non-crossing diagrams are obtained by pinching the blue regions that identify two or more indices. Furthermore, non-cactus diagrams (a7) can be identified with the crossing partition (c7).

\section{Consistency of ETH}
\label{app:consistency}
\begin{figure}[t]
	\centering
	\includegraphics[width=1 \linewidth]{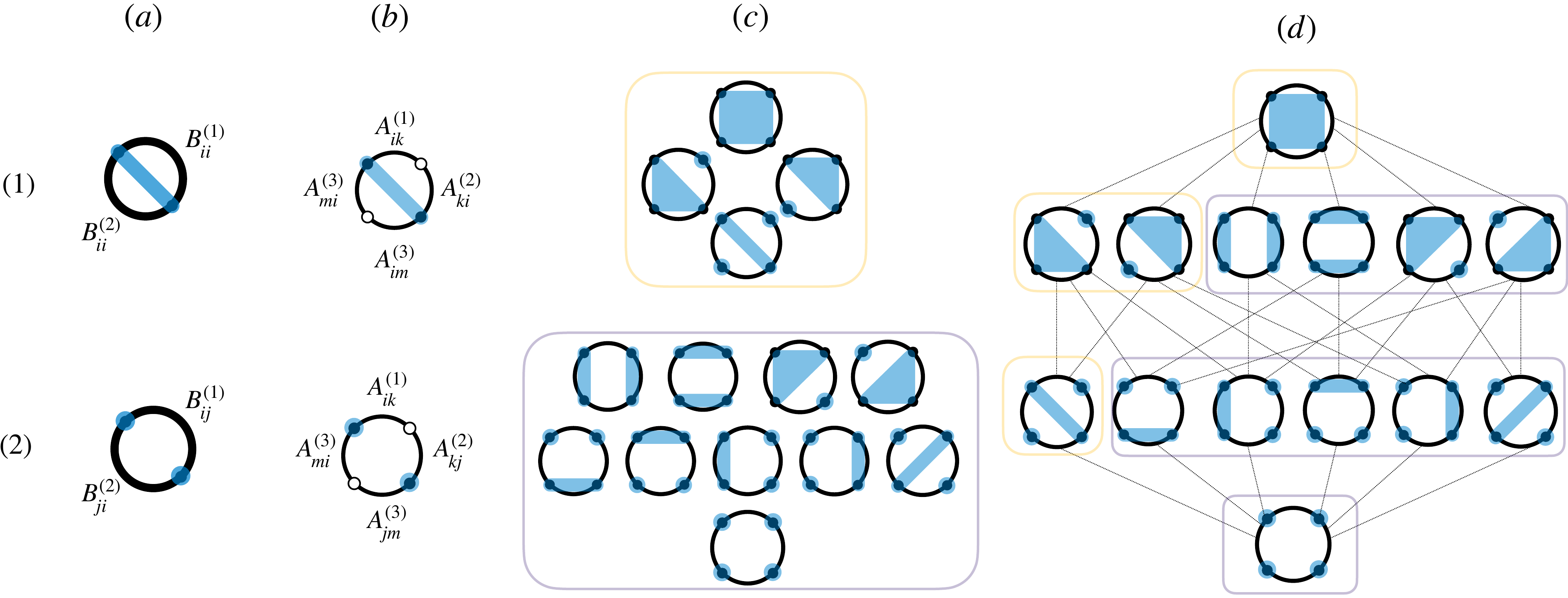}
	\caption{ETH consistency under multiplication and non-crossing partitions of Free Probability. If the operators $A$ obeys ETH, what can we say about $B = A^2$?  The lattice non-crossing partitions associated to $A^4$ in (d) can be divided in two sublattices in (c). Once summed over, this yield the non-crossing partition of the operators $B$ (a, b).   
	}
	\label{fig:consistency}
\end{figure}

Years ago, Srednicki asked the natural question about the consistency of ETH under multiplication \cite{srednicki1996thermal}. Because local functions of an observable share the `typicality' properties assumed by the observable itself, this is quite natural. However, it is instructive to see this within the framework of Free Probability.
Let us see this with an example: suppose we have $B^1=A^1 A^2$, $B^2=A^3 A^4$. If $A^1, \, A^2,\, A^3, \, A^4$ obey ETH,
can we show that $B^1,\, B^2$ automatically do too? Free Probability diagrams tell us how to restate this question in an illuminating way, as illustrated in Fig.\ref{fig:consistency}.

We draw the lattice of all dual non-crossing diagrams of the $A's$ (Fig.\ref{fig:consistency}d). With dual partition, we mean that points belonging to the same block are identified and one shall think about the blue lines as $\delta$ functions, see Section above. Here, we represent it using the \emph{lattice} structure of non-crossing partitions, see e.g. Ref.\cite{simion2000noncrossing}.  
An alternative way to obtain these non-crossing partitions is to proceed to start from the $B's$ in two steps: we first draw the non-crossing partitions for $B^{(1)}$ and $B^{(2)}$ (Fig.\ref{fig:consistency}a), where the blue lines identify the indices in the products $B^{(1)}_{ij} B^{(2)}_{ji}$. Then, every $B$ is written as a product of two $A$'s (Fig.\ref{fig:consistency}b), where the hidden index is represented as an empty dot. 
Each one of these diagrams, yields a sublattice of diagrams with all new possible lines added (Fig.\ref{fig:consistency}c), {\em excluding the ones that link two indices of the $B$'s, because they will be counted in another sublattice.} 
All these diagrams constitute a complete set of sums associated with the products of the four $A$'s, i.e. (Fig.\ref{fig:consistency}d). Just by inspection of each sublattice, we recognize the development 
of the products of two $A$'s, as for example in $ \sum_{j \neq i}A^1_{ij}A^2_{ji}+\sum_{j = i}A^1_{ij}A^2_{ji} = \sum_{j}A^1_{ij}A^2_{ji}=B_{ii}$. \\
Correspondingly, one can start from the $A's$, and -- by proceeding in the other sense -- reconstruct the non-crossing partitions of the $B's$.


The merit of this construction is that it relies exclusively on the bookkeeping of Free Cumulants.

\section{Free cumulants within ETH }
\label{app:freETH}
In this Section we compute free cumulants of Eq.\ref{free_cumuETH} of the main text within the ETH ansatz [cf. Eq.\ref{ETHq} of the main text]. One has
\begin{align}
    k_q^{\beta}(\vec t) & = \frac 1Z \sum_{i_1 \neq i_2 \dots \neq i_1 }e^{-\beta E_{i_1}} e^{i t_1(E_{i_1}-E_{i_2}) +  t_2(E_{i_2}-E_{i_3}) + \dots  t_{q-1}(E_{i_{q-1}}-E_{i_q})} 
    {A_{i_1i_2}A_{i_2 i_3} \dots A_{i_{q}i_1}} 
    \\ & =
    \frac 1Z \sum_{i_1 \neq i_2 \dots \neq i_1 }e^{-\beta E_{i_1}} e^{i \vec t \cdot \vec \omega} e^{-(q-1) S(E^+)}  F^{(q)}_{E^+}(\vec \omega) \\
    & =
    \frac 1Z \int dE_1 \dots dE_q e^{-\beta E_1 } e^{i \vec t \cdot \vec \omega} e^{S(E_1) + \dots S(E_q) - (q-1) S(E^+)}\, F^{(q)}_{E^+}(\vec \omega)
\end{align}
where from the first to the second line we have used $\vec 
\omega = (\omega_{i_1i_2}, \dots, \omega_{i_{q-1}i_q})$ with $\omega_{ij}=E_i-E_j$ and substituted the ETH ansatz and from the second to the third we have exchanged the summation with the integral $\sum _{i_1}\to \int dE_1 e^{S(E_1)}$. 
We can thus Taylor expand the entropies around energy $E^+$ as
\begin{equation}
    S(E_i) = S(E^+ + (E_i-E^+)) = S(E^+) + S'(E^+)(E_i - E^+) + \frac 12 S''(E^+)(E_i - E^+)^2 + \dots \ .
\end{equation}
Then, by summing over all the energies one obtains
\begin{equation}
    \label{eq_expEntro}
    \sum_{i=1}^q S(E_i) = q S(E^+) +  S''(E^+) \sum_i (E_i - E^+)^2 + \dots \ ,
\end{equation} 
where the linear term in $E_i - E^+$ vanishes (due to $E^+ = (E_1 + E_2 + \dots E_q)/q$), while the quadratic term is subleading to the thermodynamic property $S''(E^+) = -\beta^2 /C$ with $C\propto N$ the heat capacity and $\beta = S'(E^+)$ the inverse temperature at energy $E^+$. Since $E_i - E^+ \propto \vec \omega$ and $F_{E^+}(\vec \omega)$ is a smooth function that decays decays fast at large frequencies, we can neglect the second term in Eq.\eqref{eq_expEntro}. The free cumulant then reads
\begin{align}
    \label{eq:22}
      k_q^\beta(\vec t) = \frac 1Z \int dE_1 e^{-\beta E_1 + S(E^+)} \int dE_2 \dots dE_{E_q} e^{i \vec t \cdot \vec \omega} F^{(q)}_{E^+}(\vec \omega)   \ .
\end{align}
We can now rewrite 
\begin{align}
      E_1 & = E^+ + (E_1 - \frac{E_1 + E_2 + \dots E_q}q) = E^+ + \frac{q-1}{q}  (E_1 - E_2) + \frac{q-2}{q} (E_2-E_3) + \dots + \frac 1q (E_{q-1}-E_q) 
      \\ & = E^+ + \vec \ell_{q} \cdot \vec \omega \ ,
\end{align}
where we have defined the ladder operator 
\begin{equation}
    \label{eq:ladderQ}
    \vec \ell_q = \left ( \frac{q-1}{q}, \frac{q-2}{q} \dots , \frac{1}{q}\right ) \ .
\end{equation}
We substitute this into Eq.\eqref{eq:22} and change integration variables $dE_1 dE_2 \dots dE_1 = dE^+ d\omega_1 d\omega_1 \dots d\omega_{q-1} $, leading to
\begin{align}
    k_q^\beta(\vec t) = \frac 1Z \int dE^+ e^{-\beta E^+ + S(E^+)}
    \int d \omega_1 \dots d\omega_{q-1} e^{i \vec t \cdot \vec \omega - \beta \vec \ell_q \cdot \vec \omega } F^{(q)}_{E^+}(\vec \omega) \ .
\end{align}
Since $F^{(q)}_{E^+}(\vec \omega)$ is a smooth function of $E^+$ of order one, we can solve the integral over $E^+$ by saddle point, which simplifies with the denominator and fixed the energy by the thermodynamic definition via $S'(E_\beta) = \beta$. This immediately leads to Eq.\ref{free_thermal} of the main text.

\section{Large $\omega$ dependence of on-shell correlations}
\label{eq:large_omega}

As shown in the previous Section, the thermal free cumulant can be written at all times in terms of an on-shell correlation functions time a thermal weight [cf. Eq.\ref{free_thermal} of the main text]. At times $t=0$ this leads to
\begin{equation}
    \label{eq:wellBehaviour}
    k_q^\beta(0) =  \int d \vec \omega \,\, F^{(q)}_{E_\beta}(\vec \omega) e^{-\beta \vec \ell_{q}\cdot \vec \omega}
    =\int d \vec \omega \,\, F^{(q)}_{E_\beta}(\vec \omega) e^{-\frac{\beta}{q} ((q-1)\omega_1 + (q-2) \omega_2 + \dots + \omega_{q-1})} \ .
\end{equation}
This is just a combination of moments of the same operator at equal times and therefore it should be well defined and finite.
This imposes constraints on the behaviour of $F^{(q)}(\vec \omega)$ at large (negative) frequencies in the direction of $\ell_q$. Furthermore, the function $F^{(q)}_{E^+}$ has $(q-1)$  symmetries:
\begin{equation}
	\label{ETH3}
	F^{(q)}_{E^+}(\omega_{i_1i_2}, \omega_{i_2i_3}, \dots, \omega_{i_{q-1} i_q}) = F^{(q)}_{E^+}(\omega_{i_1i_2}', \omega_{i_2i_3}', \dots, \omega_{i_{q-1} i_q}') , 
\end{equation}
where $(\omega_{i_1i_2}', \omega_{i_2i_3}', \dots, \omega_{i_{q-1} i_q}') $ is obtained by putting $\overline \omega = \omega_{i_1i_2},+\omega_{i_2i_3}+ \dots + \omega_{i_{q-1} i_q}$,  permuting cyclically the set 
$(\omega_{i_1i_2}, \omega_{i_2i_3}, \dots, \omega_{i_{q-1} i_q},-\overline{\omega}) $,  omitting the new last term. This, together with Eq.\eqref{eq:wellBehaviour}, implies that the smooth function must fall at large frequencies in all directions at least as
\begin{equation}
F^{(q)}_{E_\beta}(\omega_1, \omega_2, \dots, \omega_{q-1})  \sim \exp(-\beta \frac{q-1}{q}|\omega_i|) \ ,
\end{equation}
which, for an exponentially decaying $F^{(q)}_{E}(\omega)\sim e^{-|\omega|/\omega^{(q)}_{max}}$, yields the bound $\omega_{max}^{(q)} \leq \frac {q-1}q \frac 1\beta$.

\section{Cumulant-moment calculation with $q=4$}
\label{sec_cumu-moment}

\begin{figure}[t]
	\centering
	\includegraphics[width=.8 \linewidth]{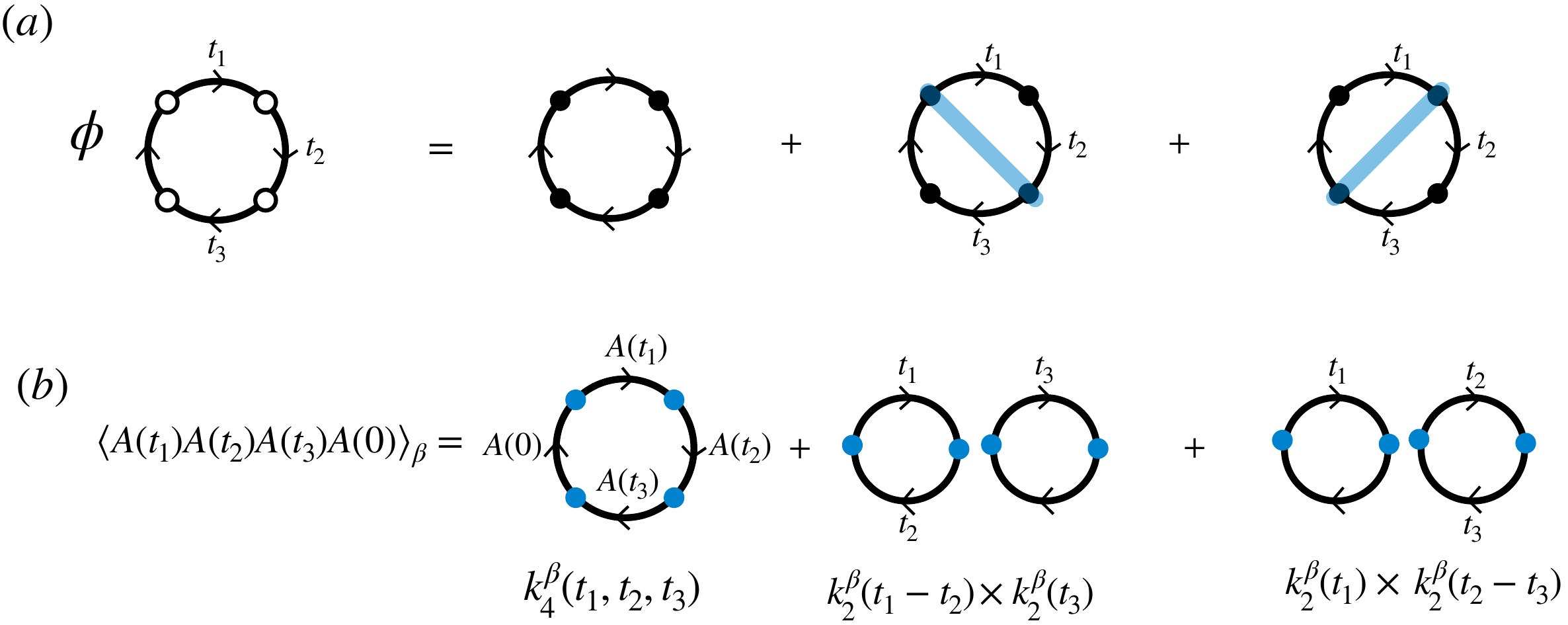}
	\label{fig:cumufree_eth}
	\caption{Relation between dual non-crossing partitions, ETH diagrams and thermal free cumulants for $q=4$ with $k_1=0$. }
\end{figure}

In this section we derive Eq.\eqref{eq_mom4} of the main text. We compute 
\begin{align}
    S_\beta^{(4)}(t_1, t_2, t_3) & = \text{Tr} \left ( \frac{e^{-\beta H}}{Z} A(t_1) A(t_2) A(t_3) A(0) \right )
    = \frac 1Z \sum_{ijkm} e^{-\beta E_i} e^{i\omega_{ij} t_1 + i\omega_{jk}t_2 + i \omega_{km} t_3} \overline{A_{ij}A_{jk}A_{km} A_{mi} }
    \\
    &  
    \label{eq:k4_app}
    =  \sum_{i\neq j \neq k\neq m} \frac{e^{-\beta E_i}}Z e^{i\omega_{ij} t_1 + i\omega_{jk}t_2 + i \omega_{km} t_3}\overline{ A_{ij}A_{jk}A_{km} A_{mi}  }
      \\
    & \quad
    + \sum_{i\neq j \neq m}  \frac{e^{-\beta E_i}}Z  e^{i\omega_{ij} (t_1 -t_2) + i \omega_{im} t_3} \overline{A_{ij}A_{ji} A_{im} A_{mi} }
     +
     \sum_{i\neq j \neq k}
    \frac{e^{-\beta E_i}}Z e^{i\omega_{ij} t_1 + i\omega_{jk}(t_2 - t_3)} \overline{A_{ij}A_{jk}A_{kj} A_{ji}} \ ,  
\end{align}
where we set $A_{ii}=0$ without loss of generality. These decompositions can be found by associating to each point on the loop and index and taking the first three non-crossing partitions. Let us now compute with ETH each term. The first coincides with the definition of free-cumulant in Eq.\eqref{free_cumuETH}. Plugging the ETH ansatz \eqref{ETHq} into Eq.\eqref{eq:k4_app}, we have
\begin{align}
    k_4^{\beta}(t_1, t_2, t_3) & = 
    \sum_{i\neq j \neq k\neq m} \frac{e^{-\beta E_i}}Z e^{i\omega_{ij} t_1 + i\omega_{jk}t_2 + i \omega_{km} t_3} F_{E^+}^{(4)}(\omega_{ij}, \omega_{jk}, \omega_{km})  e^{-3S(E^+)}
    \\
    & = 
    \frac 1Z \int dE_1 dE_2 dE_3 dE_4 e^{S(E_1) + S(E_2) + S(E_3) + S(E_4) -3S(E^+)}
    \, {e^{-\beta E_1}} e^{i\omega_{12} t_1 + i\omega_{23}t_2 + i \omega_{34} t_3} F_{E^+}^{(4)}(\omega_{12}, \omega_{23}, \omega_{34}) 
    \\
    & = \frac 1Z \int dE^+ {e^{S(E^+)-\beta E^+}}\int d\omega_{12}d\omega_{23} d\omega_{34} e^{i \vec \omega \cdot \vec t} e^{-\beta \vec \omega \cdot \vec \ell_{4}} 
    F_{E^+}^{(4)}(\omega_{12}, \omega_{23}, \omega_{34}) 
\end{align}
where from the first to the second line we have substituted summations with integrals $\sum_i = \int dE_1 e^{S(E_1)}$. From the second to the third we have expanded the individual entropies around the average one, i.e.
\begin{equation}
    S(E_i) = S (E^+ + (E_i-E^+)) = S(E^+)  + S' (E_i-E^+)  + \mathcal O(\omega^2 S'')  \ .
\end{equation}
When we sum over all of them, the term proportional to $S'$ vanishes. We have also re-written 
\begin{equation}
    E_1 = E^+ + \frac 34 \omega_{12}  + \frac 24 \omega_{23} + \frac 14 \omega_{14} 
    = E^+ + \vec \omega \cdot \vec \ell_{4}\ ,
\end{equation}
where $\ell_4$ is defined in Eq.\eqref{eq:ladderQ} and changed integration variables $dE_1 dE_2 dE_3 dE_4 = dE^+ d\omega_{12} d\omega_{23} d\omega_{34}$. We thus perform the saddle point integral over $E^+$, whose solution fixes the $E_\beta$ energy from the canonical definition $S'(E_\beta) = \beta$. This leads exactly to the result in Eq.\eqref{free_thermal} for $q=4$. Let us now compute the other term as 
\begin{align}
 \sum_{i\neq j \neq k} &\frac {e^{-\beta E_i }}Z \,\, e^{i\omega_{ij}(t_1-t_2)+ i\omega_{ik}t_3} e^{-S(E_{ij})-S(E_{jk})} F^{(2)}_{E_{ij}}(\omega_{ij})F^{(2)}_{E_{ik}}(\omega_{ik}) 
    \\
    & = 
    \frac 1Z \int dE_1  {e^{S(E_1)-\beta E_1} }
   \left (
   \int dE_2 e^{i\omega_{12}(t_1-t_2)}e^{S(E_2)-S(E_{12})}  F^{(2)}_{E_{12}}(\omega_{12})
   \right )
    \left ( \int dE_3 e^{i\omega_{13}t_3}e^{S(E_3)-S(E_{13})}  F^{(2)}_{E_{13}}(\omega_{13}) 
    \right )\\
    & =
     \left (
   \int d\omega_{12} e^{i\omega_{12}(t_1-t_2)}e^{-\beta \frac 12 \omega_{12} }  F^{(2)}_{E_{\beta}}(\omega_{12})
   \right )
    \left ( \int d\omega_{13} e^{i\omega_{13}t_3}e^{-\beta \frac 12 \omega_{13} }  F^{(2)}_{E_{\beta}}(\omega_{13}) 
    \right ) = k_2^\beta(t_1-t_2) k_2^\beta(t_3) \ .
\end{align}
Where from the second to the third line we have expanded the entropies
\begin{equation}
\label{eq:expEntro}
S(E_{2}) - S(E_{12}) = S(E_1 - \omega_{12}) - S(E_1 - \omega_{12}/2) = -S'(E_1)\frac{ \omega_{12}}{2} + \mathcal O(\frac{\omega_{21}^2}N)  \ ,  
\end{equation}
 solved the integral over $E_1$ again by saddle point and changed the integration from $E_{2/3}\to\omega_{12/3}$. We have thus found the nice factorization between the free cumulants illustrated pictorially in Fig.\ref{fig:cumufree_eth}.
We now evaluate the last term 
\begin{align}
    \sum_{i\neq j \neq k} &
    \frac{e^{-\beta E_i}}Z e^{-i\omega_{ji} t_1 + i\omega_{jk}(t_2 - t_3)} e^{- S(E_{jk}) - S(E_{ij})} F^{(2)}_{E_{jk}}(\omega_{jk}) F^{(2)}_{E_{ij}}(\omega_{ij}) 
    \\
     & = 
    \frac 1Z \int dE_2  {e^{S(E_2)} }
   \left (
   \int dE_1 e^{-i\omega_{21}t_1}e^{-\beta E_1 + S(E_1)-S(E_{12})}  F^{(2)}_{E_{12}}(\omega_{21})
   \right )
    \left ( \int dE_3 e^{i\omega_{23}(t_2-t_3)}e^{S(E_3)-S(E_{23})}  F^{(2)}_{E_{23}}(\omega_{23}) 
    \right )\\
    & = \frac 1Z \int dE_2  {e^{S(E_2)-\beta E_2} }
   \left (
   \int dE_1 e^{-i\omega_{21}t_1}e^{ \beta \omega_{21}/2}  F^{(2)}_{E_{12}}(\omega_{12})
   \right )
    \left ( \int dE_3 e^{i\omega_{23}(t_2-t_3)}e^{-  \beta \omega_{23}/2}  F^{(2)}_{E_{23}}(\omega_{23}) 
    \right )
    \\ & = 
     \left (
   \int d{\omega_{12}} e^{i\omega_{12}t_1}e^{- \beta \omega_{12}/2}  F^{(2)}_{E_{\beta}}(\omega_{12})
   \right )
    \left ( \int d\omega_{23} e^{i\omega_{23}(t_2-t_3)}e^{-  \beta \omega_{23}/2}  F^{(2)}_{E_{\beta}}(\omega_{23})\right )  = k^\beta(t_1) k^\beta(t_2-t_3)
\end{align}
where from the second to the third line we have expanded the entropies as in Eq.\eqref{eq:expEntro} and re-written the thermal weight as $e^{-\beta E_1 } = e^{-\beta E_2 +\beta\omega_{21}}$. We can thus integrate over $E_2$ by saddle point and obtain the result on the fourth line.

Putting it all together, we have
\begin{equation}
    S_\beta^{(4)}(t_1, t_2, t_3) = k^\beta_4(t_1, t_2, t_3) + k^\beta_2(t_1-t_2) k^\beta_2(t_3) + k^\beta_2(t_1) k^\beta_2(t_2-t_3)  \ ,
\end{equation}
which is exactly the (free) cumulant-moment formula in Eq.\eqref{eq_mom4}.

\end{document}